# Towards an ontology of state actors in cyberspace

Giacomo De Colle[1,2] [0000-0002-3600-6506]

[1] University at Buffalo, Buffalo NY 14260, USA
[1] National Center for Ontological Research, NY, USA
gdecolle@buffalo.edu

**Abstract.** To improve cyber threat analysis practices in cybersecurity, I present a plan to build a formal ontological representation of state actors in cyberspace and of cyber operations. I argue that modelling these phenomena via ontologies allows for coherent integration of data coming from diverse sources, automated reasoning over such data, as well as intelligence extraction and reuse from and of them. Existing ontological tools in cybersecurity can be ameliorated by connecting them to neighboring domains such as law, regulations, governmental institutions, and documents. In this paper, I propose metrics to evaluate currently existing ontological tools to create formal representations in the cybersecurity domain, and I provide a plan to develop and extend them when they are lacking.

**Keywords:** Ontologies, Cybersecurity, Cyberspace, Cyber Threat Intelligence, Data Sovereignty, Basic Formal Ontology, Common Core Ontologies.

## 1    Introduction and motivation

Cyberspace, provisionally understood as the aggregate of computing artifacts, the information they process and the connections between such artifacts, is the source of an immense and variegated body of data. Analyzing this data is at the heart of multiple disciplines, including those that are broadly located in, or related to, the field of cybersecurity, such as digital forensics, cyber threat intelligence, network analysis and incident response. Coherent integration of cyberspace data with data from neighboring fields is nevertheless difficult without a shared semantical framework such as the one that can be specified by a commonly adopted ontology. An ontology able to represent cyberspace and the operations that take place in cyberspace would allow for data sharing between different organizations, thus enabling the breaking of data silos and supporting querying, reasoning, and analysis of large bodies of data coming from different sources.

Take as an example the following case, illustrated in a simplified way in Fig. 1. An unknown device is starting a TCP handshake process, thus asking to access a certain (part of) a website containing healthcare data. This process is recorded in logs which contain information about data packets shared between different devices, and that convey information about, for example, the IP address of the device starting the TCP request. Can we automatically detect whether the request was warranted, or rather a potential attack coming from a malicious actor? In order to do so, we have to identify the type of information that the device was trying to access, as well as whether the device itself has features that are suspicious. This is possible only if we connect information from at least three different sources: the cyber operation itself (the TCP request), the type of data accessed and what it refers to (the patient and their health history), and security and criminal data, for example whether the device owner is listed in a watchlist.

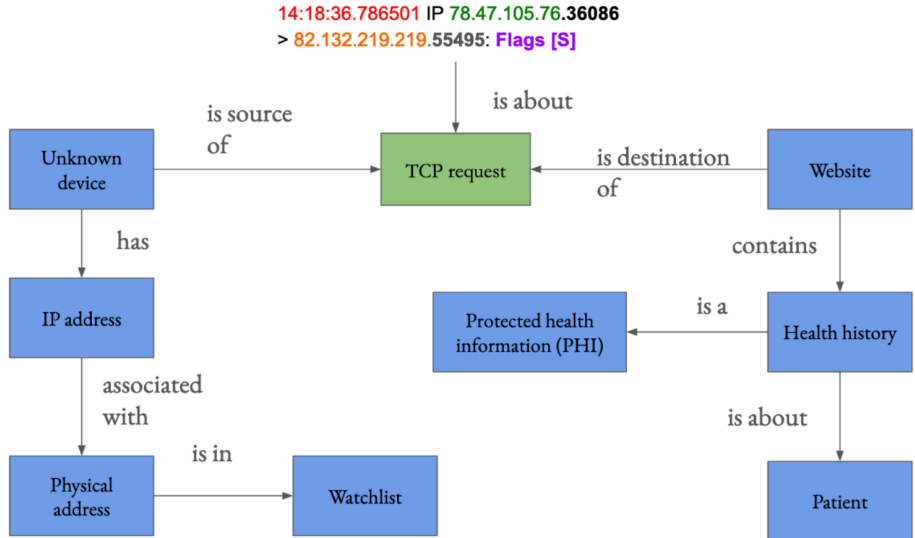

**Fig. 1:** simplified representation of a use case for an ontology of cybersecurity. Green represents BFO:occurents, blue represents BFO:continuants. Data packet information refers to different entities involved in the TCP request.

Connecting the three data sources is of course possible through direct manual intervention, but such an approach is not scalable given that it requires an exponential growth of mappings the more data sources are added (see [1] for a discussion of a similar issue in the representation of occupation data). On the other hand, an ontology structuring the data coming from these different sources would allow for its coherent integration and execution of federated queries, for example in the shape of SPARQL queries, to single-handedly identify whether the request represented in Fig. 1 is a potential cyberattack. Using semantic web technologies such as ontologies, SPARQL queries and reasoners is especially crucial when the type of information we are trying to extract from this body of data is interdisciplinary, as in the case of cyberspace information being connected with the social, legal and political domain. This is can be often the case with cyber threat intelligence practices, which have to formalize complex scenarios involving state actors, persistent threats, their values and risk assessments, as well as their motives.

In the rest of this paper, I will present a plan to use ontological resources for the modelling and formal representation of cyberspace, with a focus on the representation of state actors in cyberspace. When these ontological resources are missing, I will discuss venues to develop them. Understanding and representing state actors in cyberspace is of crucial importance as they effectively act as central units to which cyber operations, rights in cyberspace and policies owe their existence. For example, data ownership and sovereignty can be provisionally understood as the rights that state or private actors have over data regarding them, they are both mediately or immediately dependent on the state, and they are central phenomena of interest for the understanding of social and political phenomena in cyberspace. The case in Fig. 1 depicts a scenario which can potentially lead to a case of breaching of one's rights over their data. But this can only be detected in a (partially) automated way if we are able to create an ontological representation of what data ownership is. In the case represented above, the fact that the device was trying to access health record data about one's own health history makes it plausible that it is a case of data ownership breach, but this cannot be inferred and classified as such unless one has axioms connecting health data to data ownership itself.

I will begin by presenting already existing ontological efforts in the field of cyber security, and I will motivate the adoption of the Basic Formal Ontology (BFO) and the Common Core Ontologies (CCO), respectively as a top- and mid-level architecture [2, 3]. I will then present more in detail the main research question and its sub questions, alongside with associated steps in representation, formalization and ontological implementation. I will conclude with a brief presentation of the first obtained results, which currently consist in the identification of a research question, a first literature review and the creation of simple ontology design patterns for the representation of information in BFO and CCO.

## 2   State of the Art

The necessity for a formal representation of the knowledge surrounding cybersecurity has been discussed by different authors [4, 5, 6]. The extensive literature review presented in [7] shows that almost 40 efforts exist in the field of ontological representations of domains related to cybersecurity, and good ontology engineering practice requires for existing resources to be reused when possible. Recall the use case presented above in Fig. 1: in order to ontologically represent such a scenario, there is need for an ontology which is able to bridge data from cyberspace with data in neighboring domains, especially those of documents, agents, intelligence operations, and social entities. This means that the desiderata that such an ontology or ontologies need to satisfy are the following: being non-parochial, which means that they are able to be employed for different use cases of cyberspace representation; as a cognate notion, being able to be used as a hub for extensions; adopting a top-level ontology, which allows for the coherent integration of data already tagged with other ontologies, in order to avoid the creation of an ontology data silos; and the presence of technical terminology, in the ontology, which is directly tied to the domain of interest, and that doesn't suffer from being too generic. These desiderata compose metrics, which will also be used as part of the evaluation for the ontology resources I will myself develop, as discussed later on in this paper.

As shown in Table 1, many of the existing projects in the field are not tied to a top-level ontology or are developed for narrow uses such as malware recognition or risk analysis. The ontologies in question are then not able to be adopted for the type of use case presented in Fig. 1, which requires an ontology that can be used for tying together disparate data sources. The Unified Cyber Ontology (UCO), which is now part of a Linux Foundation project, was recently mapped into top-level ontologies such as BFO, but this mapping is only partial [5]. CRATELO, which adopts DOLCE as a top-level architecture [6], and the Cyber ontology, currently developed as part of an IEEE initiative, which adopts BFO and CCO as top- and mid-level architectures [8, 9], are the two projects which better merge a top-level framework with an eye towards cybersecurity. Finally, the D3FEND ontology [10], developed by MITRE, has emerged as a foundational effort in structuring cybersecurity knowledge graphs and can benefit from integration with ATT&CK, as well as one of the most promising efforts in creating a technical foundation for an ontology of cybersecurity.

Many of the use cases for ontologies representing cyberspace will interact, directly or indirectly, with neighboring areas such as intelligence analysis, directives and legal documents, defense, and counterterrorism. CCO and BFO have been recently adopted as baseline standard ontologies by the U.S. intelligence and defense communities [11]. Moreover, CCO and other projects in the BFO community already provide a baseline for representing many of the entities and phenomena related with cyberspace. Of particular interest for the purpose of this presentation are information artifacts [12], software [13], military operations [14], intelligence analysis [15], counterterrorism [16], and agents [3]. As such, BFO and CCO appear to be the privileged starting points for developing such ontological representations.

Notable efforts neighboring the ontology field are ATT&CK and D3FEND, vocabularies developed by MITRE that respectively document cyberattack and cyberdefense techniques and which are extremely valuable as data and terminology sources for the cybersecurity community. The terms included in the two vocabularies will provide a guideline for which technical terms need to be introduced in an ontology aiming to represent cyberspace.

**Table 1.** Existing ontologies of cyberspace and cybersecurity, evaluated alongside different metrics on a scale from 0 to 2, where 0 represents an ontology which doesn't satisfy the criterion, 1 represents an ontology which partially satisfies the criterion, and 2 represents an ontology which fully satisfies the criterion.

| Ontology | Non-parochial | Uses top-level | Technical | Hub |
|---|---|---|---|---|
| CRATELO | 2 | 2 | 1 | 2 |
| Cyber Ontology | 2 | 2 | 1 | 2 |
| UCO | 2 | 1 | 2 | 2 |
| D3FEND | 2 | 0 | 2 | 2 |
| COoVR, MALOnt, etc. | 1 | 0-1 | 1 | 0 |

## 3   Problem Statement and Contributions

The literature review presented in the section above identified CCO and BFO as a starting ontological ecosystem for developing an ontological representation of cyberspace and state actors. The two ontology projects which are technically better suited to represent cyberspace entities are on the other hand D3FEND and UCO. Leveraging already existing resources like the Cyber Ontology and IAO will allow for coherent mapping and integration of already existing projects like D3FEND and UCO to the BFO and CCO environment. The main research question that the project presented here addresses is whether ontologies can be used to enhance data sharing and analysis practices for cyber threat intelligence. More precisely, the project will focus on developing ontological resources to support cyber threat intelligence-enabled incident response [17, 18, 19]. To fully create an ontological basis to represent such phenomena, fields neighboring to cybersecurity also have to be explored and formally represented. Specifically, risk, values, states, rights and policies. In order to correctly identify and formally represent them, the study of texts in security studies, ethics of information and cyber warfare and risk assessment has been initiated [20, 21, 22, 23].

The metrics presented in the previous section are a first attempt at measuring success conditions for this kind of ontological application. This main research question breaks naturally down in two sub-questions. The first, regarding the ontological representation, formalization and implementation, is whether it is possible to ontologically represent cyberspace and the state and private actors that inhabit it, alongside with the laws and regulations that permeate it, the values and objectives they follow, the cyber operations they initiate, how they respond to each other, etc. Such a representation would be able to get high scores in all the metrics introduced in section 2, as well as to be able to be expanded into neighboring domains and reusable for related practices. The second, which regards the operationalization of ontologies, is whether this ontological representation can be used to support cyber threat analysis practices, as well as furnish data for the training and validation of mathematical models, simulations and AI applications.

The two main sub questions can furthermore be scaffolded into multiple and more narrow objectives. A first set of questions breaks down the first question on building an ontological representation and correspond to the first steps taken in top-down ontology engineering [24]. For what purely regards representing cyberspace and cyber operations, these questions include, for example, "what is an act of information processing?", "what is an act of information sharing over a network?", "what is a malicious actor, and what is a cyberattack?". For what regards the interplay of state actors and cyberspace, these questions include, for example "what is digital sovereignty?", "what is data ownership?", "what is an act of cyber warfare?", "what is an act of legal compliance in cyberspace?", "what is a state community, and what are its values affected by cyberspace interactions?". Answering these questions from an ontological perspective will give birth to definitions, axioms and design patterns used in ontology development.

The second set of questions breaks down the operational part of the ontology engineering practice, thus corresponding to the data-driven approach taken in bottom-up ontology development [25]. These questions will identify the applications that the ontology engineering steps undertaken or identified in other parts of the project can have. Some of these questions include, for example, "what are the formats and types that cyberspace data takes?", "what are the formats and types of data used for cyber threat intelligence purposes?", "how can cyber-attacks, related regulations and rights be represented in knowledge graph format, so that data coming from different sources can be queried for identification of these phenomena?", "what are the types of intelligence and data analysis tasks that need to be automated and integrated with data regarding threat intelligence?". Answering the second set of questions will provide the project with a more precise understanding of the type of data to be structured, and of the competency questions that such data can be used to answer. This second step will furthermore be expanded by studying and identifying the mathematical and AI models used in simulation, predictions and decision-making support in the field of cyber threat analysis [26]. The lack of data connecting the political and cyber domain is a severe problem in the creation, testing and training of these models [20], which the creation of a knowledge graph including cyberspace and state actors data can remedy.

## 4      Research Methodology and Approach

The first step of the project will involve identifying with more precision the specific type of phenomena of interest surrounding cyber threat analysis and their relationship with neighboring fields, including digital sovereignty, cyber warfare, political actors and values [20, 21, 22, 23, 24, 27, 28, 29]. Once these regulations and related phenomena are identified, it will be possible to narrow down what type of cybersecurity terms need to be used in order to identify actions that in compliance or breaking of such regulations. For example, what kind of cyberattack constitutes an act of violation of the digital sovereignty of a certain country?

Answering these types of questions within concrete existing regulations and cyber security terminologies will allow for the creation of a first set of competency questions, that will be subsequently made more precise the more ontological terminology is identified to specify them. These competency questions will eventually take the form of SPARQL queries and will be one of the evaluation tools used to assess the success or failure of the project. This first step allows for the creation of a list of *desiderata* that ontologies need to be able to satisfy, as well as a list of core terms coming from the field of law, political science and cybersecurity that are required for the success of an ontological project.

The next step involves the review of already existing ontology projects and frameworks, as well as the study of already existing neighboring non-ontological projects such as vocabularies and taxonomies. For example, MITRE's ATT&CK is one of the main resources used to represent knowledge of cybersecurity experts and effectively constitutes a well-developed knowledge basis. Other similar projects to be properly investigated include the Structured Threat Information eXpression (STIX), developed by MITRE for the DHS. The resources thus identified, alongside with the ontologies described in section 2, will then be evaluated for their capability of representing the competency questions and terms identified in the previous step. Moreover, this step will involve identifying the shape taken by data in these fields, so that the ontology can be properly structured in such a way that it mirrors it.

Once these competency questions are identified, proper development of ontology resources can begin. As previously discussed in this paper, D3FEND and UCO on the one hand, and on the other cognate ontologies in CCO and the BFO community such as the Cyber Ontology, the Information Artifact Ontology (IAO), the Agent ontology, the Information Entity Ontology (IEO), and the Geospatial ontology act as collective starting ontological basis for the development of my efforts. The method adopted in the process of ontology engineering will follow the two types of questions introduced in section 3 of this paper, thus effectively merging bottom-up and top-down ontology development in an exercise of the so-called middle-in ontology development strategy [30]. In this way, the ontological resources created can extend from the top- and mid-level layers of BFO and CCO, while also being developed with an eye to the structure of data that needs to be integrated by ontologies and to the competency questions that domain experts need to answer, as identified during previous steps of this process.

## 5   Evaluation Plan

The different steps described in the previous section can be evaluated in the course of the project in different ways. Recall that step 1 involves the identification of the notions in legal and political science which are needed in order to identify violations of state and private actors' rights over their data. This can be tested by creating definitions that are built and confronted not only with other ontologists in the community, but also with domain experts in both fields of cybersecurity and legal and political science. Success in step 1 is represented by the identification of core terms and notions in the respective fields and by the creation of first tentative definitions for them. Step 2 involves the formalization of competency questions that use terms and notions identified in the previous step. These competency questions will act as primary benchmark and use case for the ontological representations created in next steps. Step 2 will also involve the identification of the type of data that the ontologies will be created for – for example, data packet logs, server logs, access logs, and legal regulation data. Success of step 2 is marked by the creation of a satisfying set of competency questions, that are checked with the domain experts and ontologists contacted in step 1.

Step 3 will include the identification of missing terms in existing ontologies, that need to be added in order for them to represent and answer the competency questions created in previous steps. This step involves the creation of terms and definitions to be added to the ontologies mentioned above and will be evaluated by the acceptance of the terms by the ontology communities to which they are proposed to. The final step of the project will involve the ingestion of instance level data, for example taken by MITRE's database, and the testing of the ontology by means of the competency questions identified in previous steps. If the queries can be successfully applied to the knowledge base thus constructed, reasoning and implicit knowledge extraction will successfully prove the quality of the ontology and its applicability. This final part of the project will include evaluation by making use of the metrics introduced in section 2. A successful step 4 will mark a final positive result for the ontology resources created, identifying previously unidentified instances of breaking of data sovereignty or ownership. The various steps of the project can each furthermore be tested by presenting results to existing ontology conferences in the field of ontologies and law, as well as ontology and cybersecurity, such as FOIS and STIDS.

## 6     Results

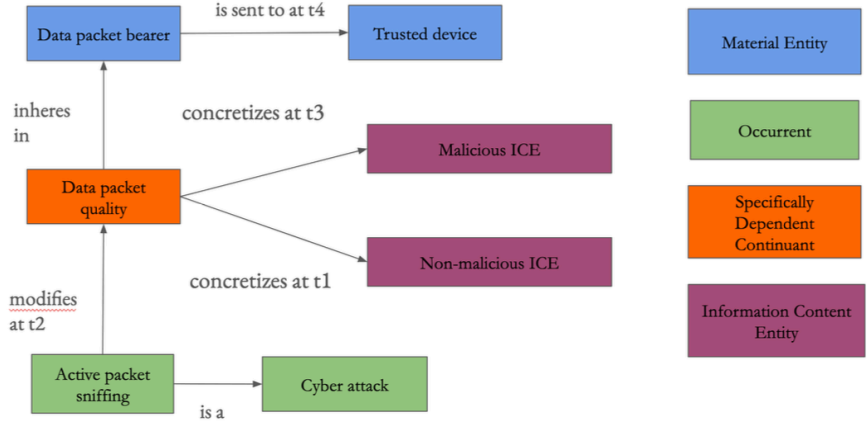

**Fig. 2:** simple design pattern for the representation of information sharing in CCO and BFO. The scenario depicted represents an active packet sniffing attack, where the content of an intercepted data packet is injected with malicious code.

The state of research is at the moment at its beginnings. Preliminary results include an extensive literature review and evaluation of the existing semantic web projects in cybersecurity, presented in section 2 of this paper. As part of these preliminary studies, I have also started identifying ontologies, data sources and already adopted terminological standards in the field, such as MITRE's ATT&CK and D3FEND, UCO, NIST recommendations, ISO standards such as ISO 27005, and STIMS. A study of foundational notions of cybersecurity has also begun, as well as contacting cybersecurity experts. One of the first objectives in the development of the project is to identify basic design patterns in CCO and BFO that can be used to represent information processing and sharing in the domain of interest of cybersecurity. Fig. 2 is an example of such first results.

## 7     Conclusions

The preliminary studies undergone so far show the need for a framework to relate efforts in ontology of cyberspace and cybersecurity with neighboring domains. To remedy this issue, I have proposed the evaluation and further development of ontological resources in cybersecurity and their interaction with ontological representations in the domain of cyber threat analysis and security studies. Such a project will achieve interoperability between heterogeneous data sources from the cybersecurity domain and the domains of security studies, information ethics and digital sovereignty. Given the difficulties proper of cybersecurity and cyber threat intelligence in data integration, this is an issue of primary concern in an era where analyzing big data for informatic vulnerabilities, sovereignty and rights infractions will exponentially develop as a focal problem.

**Acknowledgments.** The author wishes to acknowledge the help gained from comments and insightful discussions with Barry Smith, John Beverley, Fabien Gandon, the members of the 2024 Ontology Engineering and Intelligence Analysis seminar at the University at Buffalo and in particular Max Farrington, and the members of the Cyber Ontology IEEE group.

**Disclosure of Interests.** The author has no competing interests to declare that are relevant to the content of this article.

# References

1. Beverley J., Smith S., Diller M., Duncan W.D., Zheng J., Judkins J.W., Hogan W.R., McGill R., Dooley D.M., & He Y. The Occupation Ontology (OccO): Building a Bridge between Global Occupational Standards. Joint Ontology Workshops. (2023)
2. Arp, R., Smith, B., & Spear, A.D. Building Ontologies with Basic Formal Ontology, MIT Press. (2015)
3. CUBRC. White Paper. An Overview of the Common Core Ontologies. (2019)
4. Maathuis C., Pieters W. & van den Berg J. Developing a Cyber Operations Computational Ontology. *Journal of Information Warfare, 17*(3), 32–49. (2018)
5. Casey, E., Barnum, S., Griffith, R., Snyder, J., Beek, H.V., & Nelson, A. The Evolution of Expressing and Exchanging Cyber-investigation Information in a Standardized Form. in: Biasiotti, M., Mifsud Bonnici, J., Cannataci, J., Turchi, F. (eds) *Handling and Exchanging Electronic Evidence Across Europe. Law, Governance and Technology Series,* vol 39. Springer, Cham. (2018)
6. Oltramari, A., Cranor, L.F., Walls, R.J., & Mcdaniel, P. Building an Ontology of Cyber Security. *Semantic Technologies for Intelligence, Defense, and Security.* (2014)
7. Martins, B. F., Serrano Gil, L. J., Reyes Román, J. F., Panach, J. I., Pastor, O., Hadad, M., & Rochwerger, B. A framework for conceptual characterization of ontologies and its application in the cybersecurity domain. *Software and Systems Modeling, 21*(4), 1437–1464. https://doi.org/10.1007/s10270-022-01013-0. (2022)
8. Donohue, B., Jensen, M., Cox, A.P., & Rudnicki, R. A common core-based cyber ontology in support of cross-domain situational awareness. *Defense + Security.* (2018)
9. IEEE Cyber Ontology Working Group, "Cyber Ontology Releases," IEEE Open Source. [Online]. Available: https://opensource.ieee.org/cyber-ontology-working-group/cyber-ontology-releases. [Accessed 25 February 2024].
10. Kaloroumakis Peter E., Smith Michael J., "Toward a Knowledge Graph of Cybersecurity Countermeasures"
11. Lori Wade and Craig Martell. *Baseline Standards for Formal Ontology within the Department of Defense and the Intelligence Community.* Chief digital and artificial intelligence officer council members intelligence community chief data officer council members, 25 Jan. 2024
12. Ceusters W. An information artifact ontology perspective on data collections and associated representational artifacts. *Stud Health Technol Inform*, pp. 68-72. PMID: 22874154. (2012)
13. Malone, J., Brown, A., Lister, A.L., Ison, J.C., Hull, D., Parkinson, H.E., & Stevens, R. The Software Ontology (SWO): a resource for reproducibility in biomedical data analysis, curation and digital preservation. *Journal of Biomedical Semantics, 5*, 25 - 25. (2014)

14. Morosoff P., Rudnicki R., Bryant J., Farrell R. & Smith B. Joint Doctrine Ontology: A Benchmark for Military Information Systems Interoperability. *Semantic Technologies for Intelligence, Defense, and Security.* (2015)
15. Mandrick B. & Smith B. Philosophical foundations of intelligence collection and analysis: a defense of ontological realism. *Intelligence and National Security, 37*, 809 - 819. (2022)
16. Mandrick, B. An Ontological Framework for Understanding the Terror-Crime Nexus. *SOF Role in Combating Transnational Organized Crime,* 147-162 (2016)
17. Brown, Rebekah, and Scott J. Roberts. Intelligence-Driven Incident Response: Outwitting the Adversary. Second edition. Sebastopol, CA: O'Reilly Media, Inc, 2023.
18. Costa-Gazcón, Valentina. Practical Threat Intelligence and Data-Driven Threat Hunting: A Hands-on Guide to Threat Hunting with the ATT&CK Framework and Open Source Tools. Birmingham Mumbai: Packt, 2021.
19. Martinez, Roberto. Incident Response with Threat Intelligence: Practical Insights into Developing an Incident Response Capability Through Intelligence-Based Threat Hunting. Birmingham: Packt Publishing, 2022.
20. Whyte C., Mazanec B., Understanding Cyber Warfare: Politics, Policy and Strategy, Routledge 2021
21. Hubbard, Douglas W, and Richard Seiersen., How to Measure Anything in Cybersecurity Risk, Wiley 2023
22. Floridi, Luciano. Information Ethics, Its Nature and Scope, in ACM SIGCAS Computers and Society 36, no. 3, 2006
23. Security Studies, an Applied Introduction, eds. Norma Rossi and Malte Riemann, Sage 2024
24. Keet C.M. An Introduction to Ontology Engineering, ch.6. (2018).
25. Keet C.M. An Introduction to Ontology Engineering, ch.7. (2018).
26. Sarker, Iqbal H. AI-Driven Cybersecurity and Threat Intelligence: Cyber Automation, Intelligent Decision-Making and Explainability, Cham: Springer Nature Switzerland, 2024.
27. Floridi, L. (2020). The Fight for Digital Sovereignty: What It Is, and Why It Matters, Especially for the EU. *Philosophy & Technology*, 33, 369 - 378.
28. Floridi L., Mariarosaraia T. The Ethics of Information Warfare, Springer. (2014).
29. Scherenberg, F.V., Hellmeier, M., & Otto, B. Data Sovereignty in Information Systems. Electron. Mark., 34, 15. (2024)
30. De Colle G., Hasanzadeh A. & Beverley J. Ontology Development Strategies and the Infectious Disease Ontology Ecosystem. *International Conference on Biomedical Ontology.* (2023).